\documentclass[doublecol, longbibliography]{epl2}
 \usepackage{amsmath,amsthm,amssymb} 
\usepackage{graphicx}
\usepackage{hyperref}
\usepackage[active]{srcltx}
\usepackage{booktabs}

\begin{document}

\def\E{{\bf E}}

\def\Q{{\bf \Theta}}

\def\D{{\bf D}}
 
\def\r{{\bf r}}

\def\dV{{\; \rm d^3}{\bf r}}

\def\curl{{{\rm curl}\; }}

\def\grad{{{\rm grad}\; }}

\def\div{{{\rm div}\; }}

\def\p{{\bf p}}

\def\rhat{\hat{\bf r}}

\author{A.C. Maggs}
\title{Multi-scale time-stepping in molecular dynamics}

 \institute{CNRS UMR7083, ESPCI Paris, PSL Research University, 10 rue
  Vauquelin, Paris, 75005, France } \date{\today} 
 {Molecular dynamics and particle methods} 
 {Applications of Monte Carlo methods}  {Computational
  methods in statistical physics and nonlinear dynamics}
\begin{abstract} {We introduce a modified molecular dynamics algorithm
    that allows one to freeze the dynamics of parts of a physical
    system, and thus concentrate the simulation effort on selected,
    central degrees of freedom. This freezing, in contrast to other
    multi-scale methods, introduces no approximations in the
    thermodynamic behaviour of the non-central variables while
    conserving the Newtonian dynamics of the region of physical
    interest. }
\end{abstract}
\maketitle

Molecular dynamics simulation is the backbone of much numerical work
in condensed matter physics and chemistry. It is formulated in terms
of the underlying forces of the physical model and allows one to study
thermodynamic and dynamic processes in microscopic detail. However, it
is also clear that the method is, in some geometries, extravagantly
wasteful. An important and obvious example occurs in biophysical
applications where a single interesting complex is studied in the
presence of a solvent: A solvated protein requires detailed simulation
of large number of water molecules---in order to correctly hydrate the
molecule, and also to separate periodic images in the simulation cell.
Such systems seem to be obvious candidates for a multi-scale
approach. Close to the protein all the dynamics are to be followed,
but at larger distances from the protein the water molecules are
 spectators, which should give rise to self-averaging in large
systems.  Nevertheless, this ``distant spectator'' must be well modeled in
order to limit finite size artefacts.

Such ideas have led to a number of interesting papers for the
formulation of multiscale simulation \cite{ralf1, ralf2} where the
spatial resolution of the solvent model changes as a function of
position in the simulation cell. An alternative to this molecular
coarse-graining is to use different integration methods in different
regions using a simple, robust large-step integrator in the far field,
coupled to more precise near field stepper,~\cite{Leimkuhler}.

Another physical system where multi-scale ideas are important is in
the modelling of deformations in solid materials. Here short-range
atomistic modeling is coupled to larger scale continuum descriptions
in order to reduce artefacts from the tiny samples that can be
comfortably simulated \cite{elasticwave,multiscalecrystal}. It is
clear that careful matching is required to best interface the
continuum and atomistic models.

The present paper introduces a complementary view on how to treat the
far-field degrees of freedom. Rather than spatially coarse-graining or introducing
continuum couplings we  introduce a hierarchy in the speed of
simulation. Central regions in the physical system advance with a
full, molecular dynamic algorithm. Regions further away are regularly
frozen so that on average time advances more slowly. However, the
whole dynamic system is assembled in such a way that detailed-balance
is obeyed; the entire simulated system thus samples the proper
equilibrated thermodynamic ensemble. Thus there is no sharp boundary
between interesting and spectator regions. All degrees of freedom are
free to relax and fluctuate.

Physical intuition tells us that this  can be potentially 
useful: In the simulation of water the dominant local process is Debye
dielectric relaxation of the orientational dipole occurring on the
scale of $1 ps$. In a long (micro-second simulation) simulation of a
bio-molecule a single, distant water molecule influences, above all, due
to the averaged, static dielectric properties which are fully
developed at this tiny time scale. Putting more computational effort
into this rotational process than is needed is clearly
inefficient. The same can be said of a distant region in a simulation
of mechanical deformation of a solid: full high statistics simulation
of distant regions do not substantially increase the accuracy of the
simulation of the near field.

We now consider the phase space dynamics of a system which has been
partially frozen: To begin, consider the motion of a single particle
with a reduced Hamiltonian $H_i$ which contains just that part of the
full Hamiltonian $H$ that relate to the degree of freedom $i$.  We can
study the evolution of the full canonical distribution
$\rho= e^{-\beta H}$ with the reduced dynamics of only $i$ with the
Poisson bracket:
\begin{equation}
  \frac{\partial \rho }{\partial t} = \{ \rho, H_i\} = -\beta e^{-\beta H} 
  \left [ 
    \frac{\partial H}{\partial r_i} \frac{\partial H_i}{\partial p_i} - 
    \frac{\partial H}{\partial p_i} \frac{\partial H_i}{\partial r_i}
  \right ] =0 \nonumber
\end{equation}
Generalising to a mixture of free and frozen particles we see that
allowing a subsystem to evolve under the total potential energy of all
the particles conserves the canonical distribution of the full
system. At first examination this seems paradoxical. However, we note
that the full phase space has $2dN$ independent directions. If we now
allow evolution of part of a system to alternate with the full
Hamiltonian dynamics this leads to motion in which certain cuts of
phase space are explored quickly, while other cuts evolve slowly. Each
possible dynamic process, however, fully conserves the desired
measure.

The main trap in the imagined algorithm is the choice of particles in
the near and far fields in a way that does not destroy detailed
balance; this is the subject of the present paper.  In the formulation
of the algorithm we have been influenced by multi-scale Monte Carlo
methods~\cite{multimc} for the simulation of simple fluids but also an
elegant and efficient formulation of Zimm dynamics \cite{ball} as a
multiscale problem. Both these methods move blocks of particles, while
freezing the rest of the system. The originality of the present paper
is the use of a molecular dynamics, rather than a Monte Carlo rule for
the individual particle motions.

We note also that the ideas behind the algorithm are similar to a
heat-bath algorithm which works by updating a small subpart of a
physical system in a frozen background. Convergence to the Gibbs
distribution requires only the use of detailed-balance in the local
moves (or perhaps the even weaker criterion of 
balance~\cite{krauth}). Let us also note that there is a long
tradition of combining molecular dynamics and Monte Carlo rules to
produce more efficient algorithms~\cite{bias, hybrid, mcmd , shadow,
  review, brown}.

We now show how to implement the method, including the crucial choice
of the partition between frozen and dynamic degrees of freedom. The
implementation is based on a modified version of the leapfrog, or
Verlet algorithm, well known for its excellent conserving properties.
We define the operators for updating the velocities of particles, the
positions of particles and for thermalising particles as follows:
\begin{align*}
  L_v (dt) : \quad v &\leftarrow v + f(r)\, dt \\
  L_r (dt) : \quad r &\leftarrow r + v\, dt \\
  L_T (dt) : \quad v &\leftarrow e^{- dt\, \eta  } v + \sqrt{{T (  1-e^{ -2 dt\, \eta}  )}} {\mathcal N}(0,1) 
\end{align*}
where, for simplicity we hide the vectorial nature of the update
rules.  $T$ denotes the temperature and $\eta$ a damping rate for
velocity correlations due to interaction with a Langevin
thermostat. The velocities, positions and forces are denoted
respectively by $v$, $r$ and $f$.  ${\mathcal N}(0,1)$ is a normally
distributed random number with unit variance. We work with unit mass
particles.

These individual steps are assembled into a time-reversible
sequence~\cite{leimtop}:
\begin{equation*}
  L(dt) = L_r(dt/2)\, L_T(dt/2) \, L_v(dt)\,   L_T(dt/2) \, L_r(dt/2)
\end{equation*}
Evolution for $M$ time steps gives the operator ${[L(dt) ]}^M$.
The half step updates can be coalesced to give the thermalised
leap-frog algorithm~\cite{tuckerman}.  This defines a map on phase
space compatible with detailed balance and the thermalised process
converges to the canonical ensemble. We now add a further action in
the dynamics. We stochastically displace (with probability $p$) a
spherical barrier of radius $\ell$: $B_\ell$ after application of the
operator $L(dt)$. This barrier, centred on the region of interest, has
a radius drawn from a probability density
\begin{align}
  P(\ell) =& 0 \quad \ell < \ell_0 \nonumber  \\
  \sim &\frac{1}{\ell^{d+1}} \quad \ell > \ell_0 
         \label{eq:prob}
\end{align}
The whole dynamic process then involves an alternation of microscopic
leap-frog evolutions and barrier events:
\begin{equation}
  \dots L(dt) B_\ell   L(dt) \dots
\end{equation}
in manner which is still explicitly time reversible.

\begin{figure}[ht]
  \includegraphics[scale=.3] {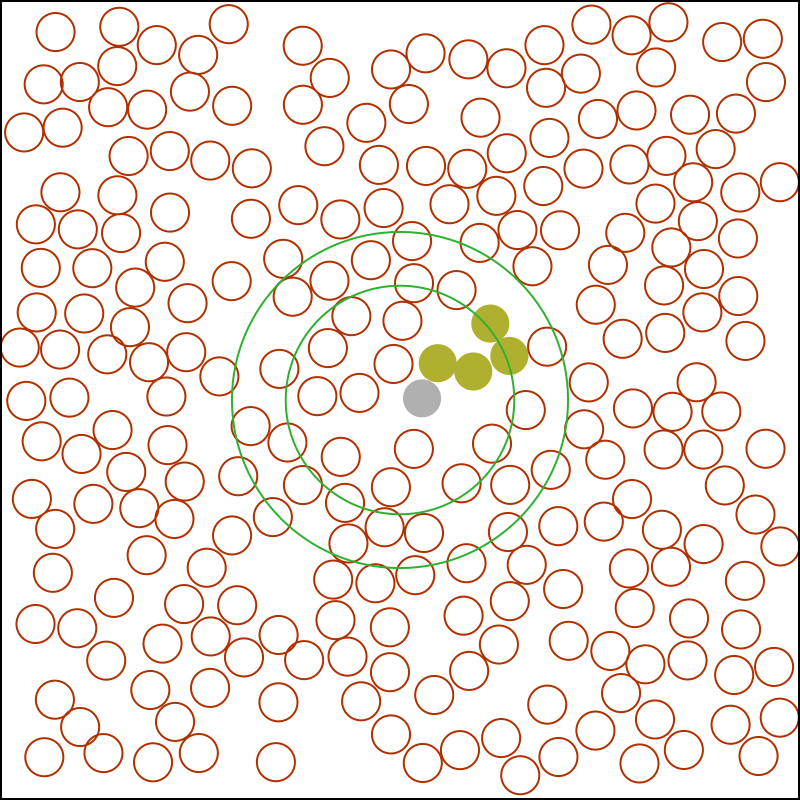}
  \caption{Snapshot of a simulation of a short polymer chain in fluid
    background. Hollow (red) particles are monomers, solid particles
    form the polymer, with a single particle tethered to the centre of
    the simulation cell. We use periodic boundary conditions.  Inner
    ring: a radius corresponding to $\ell_0 =2.5 \sigma$.  Large outer
    ring: the specific choice of the barrier radius $\ell$ at this
    moment in the simulation.  Particles beyond the outer ring are
    frozen, particles within the inner ring are always mobile. Here
    the chain has been split between inner and outer regions so only
    part of it is dynamic at all times. Simulation cell size
    $L=20\sigma$, 240 particles.}
  \label{fig:snap}
\end{figure}

This barrier is selected with rather special properties. It
does not interact with any particle in the system; the insertion has
zero energy cost. However, the barrier changes the dynamics of the
system strongly. All particles exterior to the barrier
Fig.~\ref{fig:snap} are frozen in position. All particles within the
barrier move in the total potential of the other interior particles
\textit{plus} the frozen exterior particles. When the radius of a
barrier increases the particles restart their motion with
\textit{exactly the same velocity} they had before being frozen.

As stated this rule still has a major defect: particles in a time step
can cross the barrier where their dynamic status becomes ambiguous. If
they are then frozen the dynamics becomes  irreversible since
forwards and backwards motions are not generated in a manner
compatible with detailed balance.  We thus add \textit{one extra
  feature} to the barrier --- \textit{it acts as a perfect mirror} to
any incident interior particle. This reflecting barrier, together with
the careful insertion of the barrier event within the half-step of the
leap-frog method give rise to a dynamic system which is explicitly
time reversible. Management of this extra dynamic process is treated
using event-driven methods within the step
$L_r(dt/2)$~\cite{leimhard}. If we take a particle which has bounced
off the barrier in the time $dt/2$ and reverse its speed it will
backtrack on its own trajectory to within machine precision.

The probability distribution for the barrier gives
$\ell_0 < \ell <\infty$ with a preference for a radius which is close
to $\ell_0$. With this choice, eq.~(\ref{eq:prob}), the total
numerical effort only increases very slowly
(logarithmically)~\cite{multimc,ball} with the total system size: yet
all particles move and relax during the simulation. As one goes away
from the region of interest, the total simulation effort decreases.

We have thus a dynamic process on phase space respecting detailed
balance, which in many ways is reminiscent of heat-bath Monte Carlo
where a selected set of variables are re-sampled in the static
background of the entire system. However, within the interior region
of radius $\ell_0$ the particles always evolve under Newtonian
dynamics. It is clear that the dynamics are not momentum
conserving---collisions on the barrier act like a confining box. The
method is thus unlikely to be useful in the case of physics dominated
by long-ranged hydrodynamic interactions.

We  applied the algorithm to the a simple two-dimensional
Lennard-Jones fluid and studied the partial structure factors of a
central region, and compared to a direct molecular dynamics simulation
of the whole system. We use an integration time step $dt=0.005$, with
damping $\eta=0.2$ in Lennard-Jones units and cut-off the interaction
at $2\sigma$.  It was in this preliminary study that we became aware
of the importance of the correct time-reversible version of the
algorithm. Insert of the barrier in a non-symmetric postion in the
standard leapfrog method led to a systematic deviation in measured
correlations.

We then performed simulation with a short (5-mer) tethered polymer in
a Lennard-Jones fluid, Fig.~\ref{fig:snap}: Such small-scale
simulation where very high statistics runs can be performed are those
that are most sensitive to errors in the formulation of the algorithm.
In the snapshot we draw two rings---the innermost corresponding to
$\ell_0=2.5 \sigma$ in Lennard-Jones units. The outer ring is the
instantaneous position of the reflecting barrier chosen from the
probability distribution eq.~\ref{eq:prob}. We see the inner ring
intercepts the chain. It is thus stochastically split between the
mobile inside and the static outside region during its
evolution. Despite this extreme perturbation to the dynamic the
conformations of the chain are unperturbed. In Table~\ref{tab:table1}
we report on the mean squared end-to-end separation, $R_e^2$ of the
5-mer as a function of the minimum barrier radius. We measure
deviations in the polymer size, and give estimates of statistical
errors.

\begin{table}[h!]
  \centering
  \caption{Estimates for the polymer size compared to average over all simulations of $R_e^2=5.2443
    $. Deviations from the mean are compatible with the estimated
    statistical errors. We used a value $p=0.01$ for the probability of a change
    in $\ell$ in each time step.}
  \label{tab:table1}
  \vskip 0.5cm
  \begin{tabular}{ccc}
    \toprule
    $\ell_0$ & deviation in $R_e^2$& statistical error \\
    \midrule
    2.5 & -0.0025 & $ 0.002$\\
    3 & -0.0004 & $ 0.002$\\
    4 & -0.0005 & $ 0.0015$\\ 
    5 & 0.0008 & $ 0.0015$\\
    6 & 0.0015 & $ 0.0015$\\
    8 & 0.0009 & $ 0.0015$\\
    10& 0.0004 &$ 0.0015$\\
    $\infty$ & 0.001 & $ 0.0015$\\
    \bottomrule
  \end{tabular}
\end{table}

We conclude that to a precision of better than 1 part in 2000 the
mean-squared extension of the polymer is independent of $\ell_0$. This
requires simulations of length $10^7$ autocorrelation times. Note,
that when we try simulating the polymer chain in small, periodic cells
we see substantial modifications of the statistics of the chain when
using cells of dimension $2 \ell_0$ with the smallest values of
$\ell_0$ in Table~1.

\begin{figure}[ht]
  \includegraphics[scale=.51] {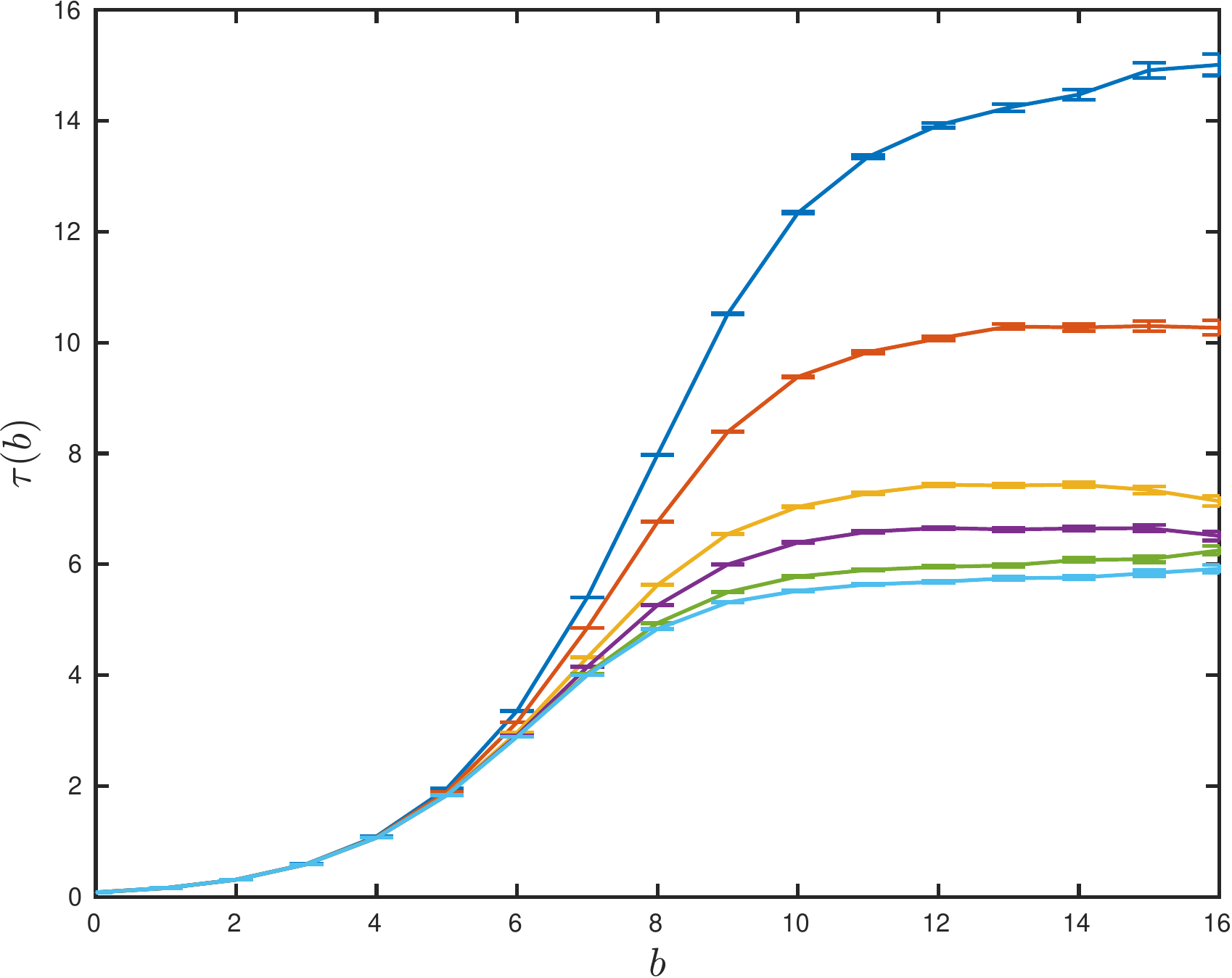}
  \caption{Blocking analysis of the autocorrelation of $R_e^2$ for a
    5-mer in a monomer background, system shown in
    Fig.~\ref{fig:snap}.  $\tau(b)$, eq.~(\ref{eq:tau}), in
    Lennard-Jones units, as a function of blocking, $b$. From top to
    bottom: $\ell_0=2.5 \sigma$, $\ell_0=3\sigma $, $\ell_0=4\sigma$,
    $\ell_0=5\sigma $, $\ell_0=8 \sigma$, full system. As a function
    of the blocking factor $\tau(b)$ varies, saturating for large $b$
    which is an estimate of the autocorrelation time.}
  \label{fig:auto}
\end{figure}

We now study the autocorrelation dynamics of the chain using a method
based on block averages~\cite{error}, Fig.~\ref{fig:auto}: A long time
series with variance $\sigma^2_v$ can be averaged into blocks of
length $t_b=t_{rec} 2^b$, which themselves have an interblock variance
$\sigma_b^2$.  $t_{rec}$ is the time in Lennard-Jones units between
recordings in the time-series. If the blocks are longer than the
integrated autocorrelation time, $\tau$, we have
$ \sigma^2_b=2 \sigma_v^2 \tau/t_b$. Thus as the block size increases
an estimate of the autocorrelation time is
\begin{equation}
  \tau(b) =\sigma_b^2t_b/(2 \sigma_v^2)
  \label{eq:tau}
\end{equation}
When we plot this ratio as a function of $b$ it should plateau to a
constant value, when the block size is longer than the autocorrelation
time.  In Fig.~\ref{fig:auto} this gives rise to a characteristic
rising curve $\tau(b)$ which encodes the autocorrelation dynamics of
the chain. The value of the large $b$ plateau is then a true estimate
of the integrated autocorrelation time. In the figure one can read-off
the approximate autocorrelation time for several values $\ell_0$. The
highest curve, corresponding to the slowest relaxation, is for
$\ell_0 =2.5 \sigma$. We see that the dynamics has been strongly
hindered due to the regular partition of the polymer between frozen
and unfrozen zones. However, all the curves for $\ell_0 \ge 5 \sigma$
are close, corresponding to the large system limit.  Table~1 again
shows that the dynamic process always generates the correct statistics
of the chain, even if the parameters are chosen in such a way to
strongly change the underlying dynamics.

We have demonstrated the possibility of embedding a full speed
molecular dynamics simulation in a larger, more slowly evolving
background in such a way that the whole system generates the canonical
ensemble of the entire system. This gives \textit{essentially perfect
  boundary conditions} for the simulation of the central region
without having to build a coarse-grained model of the outside
world, and without needing to couple a continuum theory.

We anticipate that such methods will be useful in the simulation of
solvated molecules or in large scale simulation of elastic materials;
in such systems much computational effort is lost in updating weakly
interacting spectator molecules, rather than concentrating the
computation effort on the specific object of interest.  For use in
situations where long-ranged interactions are important, as is the
case with water one needs methods of evaluating efficiently the
interactions in the presence of many frozen particles. One possibility
is the use of a dynamic, propagative algorithm for the electrostatic
field as has been presented
in~\cite{localmd,nanoscale,localmc,dynamics}.

In this paper we used a specific distribution for the barrier
dimension eq.~(\ref{eq:prob}) which has been shown useful in Monte
Carlo applications, but it is probable that the optimal distribution
of work between near and far fields is problem dependent.

\acknowledgments We thank Ralf Everaers and Michael Schindler for
extensive discussions on the formulation of multi-scale simulation
methods.

\bibliographystyle{eplbib} \bibliography{var}

\end{document}